\documentstyle[12pt]{article}
\begin{document}
\textheight 24.cm
\textwidth 16cm
\oddsidemargin 0in
\topmargin 0in
%\noindent
%{\bf {\em To appear in Journal of Physics: Condensed Matter}}
%\vskip .1in
\begin{center}
\large {\bf Hubbard model with bond-charge interaction
on a triangular lattice: a renormalization group study}
\end{center}
\vskip .03in
\begin{center}
{\em  Bibhas Bhattacharyya$^\dag$ and Shreekantha Sil$^\ddag$}
\vskip .02in
$^\dag$ Department of Physics, Scottish Church College, 1 \& 3 Urquhart Square,
Calcutta 700 006, India \\
$^\ddag$ Institut f$\ddot{\rm u}$r Theoretische Physik, Universit$\ddot{\rm
a}$t zu K$\ddot{\rm o}$ln, Z$\ddot{\rm u}$lpicher Strasse 77, D-50937,
K$\ddot{\rm o}$ln, Germany
\end{center}

%\vskip .5cm
\vskip .03in

%\newpage

\centerline {\bf Abstract}

\noindent
{\small
We have studied the Hubbard model with bond-charge interaction on a
triangular lattice for a half-filled band. At the point of particle-hole
%symmetry we obtain exact solutions in two wide regimes of the 
symmetry the model could be analyzed in detail in two opposite regimes of the
parameter space. Using a real space renormalization group we calculate the
ground state energy and the local moment over the whole parameter space.
The RG results obey the exact results in the respective limits. In the intermediate
region of the parameter space the RG results clearly show the effects
of the non-bipartite geometry of the lattice as well as the absence
of symmetry in the reversal of the sign of the hopping matrix element.}

%\newpage
\vskip .04in
\noindent
{\bf I. Introduction}
\vskip .03in

	In spite of an extensive effort over the last few years the problem of
 electronic correlation in low dimensional systems remains yet to
be clearly understood. Models of interacting electrons  are difficult
to handle in one and two dimensions owing to strong fluctuations.
While there are a few exact solutions in one dimension (1-D) [1] and in
infinite dimensions [2]  the
situation is worse in two dimensions (2-D) . Standard techniques like
mean field
 approximation or perturbative calculations are of very limited use in treating 
the intermediate to strong correlation in low dimensional systems which are
typically dominated by strong fluctuations. Numerical simulations
and exact diagonalizations are also limited by small cluster size
because the dimension of the Hilbert space soon becomes too large to
be handled as one goes from one to two dimensions. Therefore,
it seems enterprising to apply an approximate real space renormalization
group (RG) known as the block RG (BRG) [3,4] in this context. 
This works reasonably well for 1-D systems [5-- 7] and over the
whole range of the coupling constants. This method employs
a truncation of the Hilbert space, retaining only a few low lying states only,
to bring out the essential ground state properties. Although the efficacy
of the truncation procedure is much more satisfactorily handled in a recently
developed RG scheme known as the density matrix RG (DMRG) [8], the latter
is yet to be developed for a truly 2-D system. The BRG has already been
applied to interacting electrons in 2-D [9] but only for bipartite lattices.
It is interesting to see how does it work in case of a non-bipartite lattice
like the triangular one. On the other hand the problem of interacting electrons
on a triangular lattice has been addressed for a long time [10-- 13] in view
of a rich phase structure including  the possibility of  frustration [14].
Particularly, $^3$He on graphite was considered to be a good example of
the Hubbard model [15] on a triangular lattice [10]. 
Recent works on the organic superconductors like $\kappa$-BEDT-TTF compounds
also investigate the phase diagrams of interacting-electron models on
anisotropic triangular lattices [16].
In the present work we
study a generalized Hubbard model with bond-charge interaction on the 2-D 
triangular lattice at half-filling by using the BRG. In Sec. II we present
the model and the renormalization scheme as suitably adapted to the present 
problem. Sec. III summarizes some of the exact solutions of this problem
in limiting cases. In Sec. IV we focus on our results obtained from the
RG and make comparisons with the exact solutions. Sec. V concludes the
present work.

\vskip .04in
\noindent
{\bf II. The Model and the RG scheme}
\vskip .03in

	The generalized Hubbard model with bond-charge interaction is defined as
follows:
\begin{eqnarray}
H&=&-t \sum_{ \langle ij \rangle ,\sigma}c^{\dag}_{i \sigma}c_{j \sigma} 
+ X \sum_{ \langle ij \rangle ,\sigma}
c^{\dag}_{i \sigma}c_{j \sigma}(n_{i -\sigma} + n_{j
-\sigma}) - \mu \sum_i n_i \nonumber \\
& &+ U \sum_i n_{i \uparrow}n_{i \downarrow} ,
\end{eqnarray}
where $c^{\dag}_{i \sigma}(c_{i \sigma})$ creates (annihilates) a
particle with spin $\sigma$ in a Wannier orbital located at the site $i$;
the corresponding number operator is $n_{i \sigma} = c^{\dag}_{i
\sigma}c_{i \sigma}$. 
 Here $n_i = n_{i \sigma} + n_{i -\sigma}$. 
Sum over $ \langle ij \rangle $ denotes contributions from distinct nearest
neighbour pairs of sites on a triangular lattice. 
 $\mu$ is the chemical potential. We focus our attention to the special
case with $X=t$ for which the present model becomes particle-hole (p-h)
symmetric even on the triangular lattice. In fact it has been shown
that for some realistic systems $X \simeq t$ [17]. Further the choice  
$X=t$ results in some simplification in implementing the RG and forces 
$\mu=U/2$ for half-filling. Moreover, at this special
value of $X$ we have some exact results for comparison.
One should remember, however,
that there is no $t \rightarrow -t $ symmetry due to the non-bipartite
nature of the lattice.

	Now, for $X=t$ and for a half-filled band ($\langle n_i \rangle=1$)
one can put $\mu=U/2$ to rewrite the Hamiltonian (2) as follows:
\begin{eqnarray}
H&=&-t \sum_{\langle ij \rangle ,\sigma}c^{\dag}_{i \sigma}c_{j \sigma}(1 - n_{i
-\sigma} - n_{j -\sigma}) + U \sum_i\left(\frac{1}{2} - n_{i
\uparrow}\right)\left(\frac{1}{2} - n_{i \downarrow}\right)\nonumber\\
& & + D \sum_i {\bf 1}_i +\frac{U}{2}N,
\end{eqnarray}
where, the constant $D=-\mu/2$ will account for the 
renormalization of the vacuum energy. 
Here we have added a constant term $NU/2$ to the Hamiltonian given by (1)
to compensate for the chemical potential term subtracted thereof. We just
keep aside this constant term for the purpose of renormalizing the parameters
in the Hamiltonian and then add up the same to the ground state energy at the
 end.  The Hamiltonian has several conserved 
quantities: total number of particles $\nu$, total spin $S$ and the 
$z$-component $S_z$ of the total spin besides the p-h symmetry  pointed out 
earlier.  We divide the 2-D lattice with $N$ sites into $N/3$ triangular blocks
of three sites each (Fig. 1). The block-Hamiltonian is then diagonalized for the
three-site block. Since we are interested in the ground state properties of the 
system we truncate the Hilbert space for the three-site blocks and retain
only four low-lying states of the block-Hamiltonian governed by the symmetries
mentioned above together with the point group symmetry of the block. 
It is easy to see that the block-Hamiltonian is block-diagonalizable in terms
of the good quantum numbers $\nu$, $S$ and $S_z$.
The states with $\{ \nu=2, S=S_z=0\}$, $\{ \nu=3,S= S_z = 1/2 \}$, 
$\{ \nu=3,S=- S_z = 1/2 \}$ and $\{ \nu=4, S=S_z=0 \}$ are considered for this
purpose. Of these the states in the first and the fourth group are connected by
the p-h symmetry while those in the second and third by spin-reversal
symmetry. However, the point group symmetry in the 2-D lattice imposes
further restriction on the choice of the states [9].
	
	To resolve this point we consider the symmetry group $C_{3v}$ [18] of
the basic triangular block. We take our eigenstates of the block-Hamiltonian to
also be the eigenstates of $R(C_{3v})$, the matrix representations of the
group elements of $C_{3v}$. While choosing the four states to be retained,
we simply take one from each of the  four groups so that all of them
lie in the same representation of $C_{3 v}$ such that the contribution to the 
ground state energy be the minimum.
This leads to the following 
RG equations which relate the renormalized parameters (primed quantities) 
to the original parameters (unprimed ones) in the Hamiltonian :
\begin{eqnarray}
 U'&=&2(E_2-E_3),\nonumber\\
 D'&=&3D + (E_2+E_3)/2,\nonumber\\
 t'&=& 2 Re[~\lambda^{\ast} (\lambda-2\lambda')~]~t~,
\end{eqnarray}
where $E_2$ and $E_3$ are the lowest eigenvalues in 
the subspaces $\{\nu=2,~S=S_z=0\}$ and $\{\nu=3,~S=S_z=1/2\}$ 
respectively; and,
\begin{eqnarray}
&&\lambda = \langle \nu=2,~S=S_z=0|c_{b \uparrow}|\nu=3,~S=S_z=\frac{1}{2} 
\rangle \nonumber \\
&&\lambda' = \langle \nu=2,~S=S_z=0|c_{b \uparrow}n_{b \downarrow}|
\nu=3,~S=S_z=\frac{1}{2} \rangle~. 
\end{eqnarray}
$\lambda^{\ast}$ is the complex conjugate of $\lambda$ and $Re$ denotes the real part.
Here the subscript $b$ refers to the site index of a {\em boundary} site of 
the block. For a triangular block, however, this could be any of the three 
sites. The factor of 2 appears in the renormalization of the hopping due to the
two connecting paths between the two neighbouring three-site blocks 
(Fig. 1). We illustrate the scheme of the RG calculations using the recursion
relations (3) in the Appendix.

	Using these recursion relations one can find out the ground state energy
$E_0$ by
$$
E_0=\lim_{n\rightarrow \infty} D^{(n)}+NU/2~,
$$
where the superscript refers to the $n$-th stage of iteration. The term
$NU/2$ in the above expression comes from the constant term added in (2).
 One could
study the recursion of any suitable operator to calculate any other physical
quantity within this approximation. For example, we calculate the local moment
$L_0$ given by
$$
L_0=\frac{3}{4}\left( n_{\uparrow} -n_{\downarrow} \right)^2
$$
The recursion relation for $L_0$ turns out to be
$$
L_0= A +B\,L'_0
$$
where $A$ and $B$ are numbers depending on the  components of the basis
states retained in course of basis-truncation. Since all the three sites
in the basic triangular block are in the same status (after adapting
the $C_{3v}$ symmetry) one can use any of them for finding the recursion of
$L_0$.

	It is interesting to note that due to the group theoretical restriction
on the choice of the truncated basis it is not possible, in general, to
select the lowest energy states from all the four groups at a time. The 
states with the lowest energies in all the four groups do not
necessarily belong to the same irreducible representation of
$C_{3v}$. If such states are retained then the matrix element $\lambda$
defined in (4) will be zero. So instead of the states with lowest energy
one targets the states belonging to the same irreducible representation.
  This problem was
not present in the 1-D cases. Therefore, it is possible here to choose
any of the possible combinations of four states (from the four groups)
compatible with the symmetry group. Again from (3) one can see that
the contribution to the ground state energy from each iteration is
$\propto (E_2+E_3)$. Then it is natural to seek for the combination
which gives the lowest value of this quantity. However, one should
be careful about the value of the renormalized hopping $t'$ thus generated,
because this in turn will seriously affect the contribution to the energy
in the subsequent iteration. We have, therefore, taken into consideration
all the possible channels permitted by the symmetry group up to three
subsequent iterations and then considered the  energetically 
best ones for the next iterations. For example, consider that  we have 
an optimum choice of five distinct channels (each with a unique set of four
states belonging to a given  irreducible representation)
to start with. In three successive iterations it gives rise to
125 ($=5\times 5\times 5$) channels out of which we retain  the best
twenty five for the next step. This is an optimized way to achieve the 
true ground state. Different possible combinations of the two- and
three-particle states are  shown in Table 1 in terms of the irreducible
representation, and corresponding energies $E_2$ and $E_3$. It is evident from
the table that the contribution $(E_2+E_3)/2$ to the constant term $D$
in the Hamiltonian (2) makes only a few combinations preferable in
constructing the RG scheme.
\vskip .02in
%\centerline {\bf Table 1}
\vskip .01in
\begin{center}
\begin{tabular}{|c|c|c|c|} \hline
Combination given by& $E_2$& $E_3$ & $E_2+E_3$ \\
irreducible representation&&& \\ \hline 
$A_1$ & $-2t-U/4$& $U/4$&$-2t$ \\ \hline 
$A_1$ & $3U/4$&$U/4$&$U$ \\ \hline
$E$&$t-U/4$     &$-3U/4$   & $t-U$ \\ \hline
$E$&$t-U/4$  &$-\sqrt{3}t+U/4 $& $(\sqrt{3}+1)t$    \\ \hline
$E$&$t-U/4$  &$\sqrt{3}t+U/4 $& $-(\sqrt{3}-1)t $   \\ \hline
$E$&$3U/4$     &$-3U/4$      & $0$    \\ \hline
$E$&$3U/4$     &$-\sqrt{3}t+U/4 $&$ \sqrt{3}t+U$    \\ \hline
$E$&$3U/4$     &$\sqrt{3}t+U/4$ & $\sqrt{3}t+U$  \\ \hline  
\end{tabular}
\end{center}
\vskip .01in
\noindent
{\bf Table 1:} Energies of the eigenstates of the block-Hamiltonian
corresponding to $\nu=2$ and $\nu=3$ are $E_2$ and $E_3$ respectively. These are tabulated for different possible combinations of states belonging to
different irreducible representations of $C_{3v}$. We adopt standard group 
theoretical notation [18] ($A_i$ for 1-D irreducible representations while
$E$ for the 2-D one).
\vskip .04in
\noindent
{\bf III. Exact Results for the Model}
\vskip .03in

Exact ground state energy could be calculated for the present model (2)
in two opposite extremal ranges of the parameter space in $U/t$. To do this
we try to match a variational upper bound $E_{\rm up}$ to the exact energy
to a lower bound $E_{\rm lo}$. To find out the upper bound $E_{\rm up}$ we choose 
a trial state $\mid \Psi_{\rm trial} \rangle$ and calculate, by variational 
principle,
$$
E_{\rm up} = \frac{\langle \Psi_{\rm trial} \mid H \mid \Psi_{\rm trial} \rangle}
{\langle \Psi_{\rm trial} \mid \Psi_{\rm trial} \rangle}~.
$$
To calculate the lower bound $E_{\rm lo}$ we break up the Hamiltonian (2)
into a sum of the cluster-Hamiltonians corresponding to the smallest
triangular clusters. These could be exactly diagonalized. 
Then, again by variational principle, $E_{\rm lo}$ is given by
$$
E_{\rm lo}= \sum_{\alpha} E_{\rm min}^{\alpha} +C
$$
where $E_{\rm min}^{\alpha}$ is the lowest eigenvalue  for the $\alpha$-th cluster
and $C$ is the constant term, if any,  appearing in the Hamiltonian.
To find out the lower bound we rewrite the Hamiltonian (1) in the following form
\begin{eqnarray}
H&=&\sum_{\alpha = 1}^{2N}\left[-\frac{t}{2} \sum_{\langle ij \rangle \epsilon \alpha,\sigma}c^{\dag}_{i \sigma}c_{j 
\sigma}\left(1 - n_{i
-\sigma} - n_{j -\sigma}\right) + \frac{U}{12} \sum_{i \epsilon \alpha}
\left( n_{i \uparrow} - n_{i \downarrow}\right)^2 \right]\nonumber\\
& & +\frac{U}{2}N~ ~, \nonumber \\
&=& \sum_{\alpha=1}^{2N} H_{\rm cluster}(\alpha)+\frac{U}{2}N ~,
\end{eqnarray}
where we have used $\mu=U/2$ and compensated for the same by adding a term
$NU/2$, $N$ being the number of lattice sites. So, here, $C=NU/2$.
In the above form of the Hamiltonian we have summed over all possible
($2N$ in number) triangular clusters $\alpha$. The fractional numbers appearing
with $t$ and $U$ are due to the over counting of the bonds (twice each) and
the sites (each being shared by six adjacent triangles). $H_{\rm cluster}
(\alpha)$ is the cluster Hamiltonian of the $\alpha$-th cluster for which
we have to find out the lowest eigenvalue $E_{\rm min}^{\alpha}$.

We categorize the results in two regimes as follows:

\noindent
I. $U>0$: For very large positive values of $U$ the system is expected to go 
over to a phase with singly occupied sites [17]. We choose the trial 
wavefunction as  
\begin{equation}
|\Psi \rangle = \prod_{i \in {\cal L}} c_{i \uparrow}^{\dag}\prod_{j \in {\cal L'}}
 c_{j \downarrow}^{\dag}~\mid \! 0 \rangle,
\end{equation}
where $\mid \!0\rangle$ is a site-vacuum and $\cal L$ and $\cal L'$ are
arbitrary disjoint sets of lattice sites, each containing $N/2$ lattice points,
which together build up the whole lattice. Using this $|\Psi\rangle$ we obtain
an upper bound $E_{\rm up}=0$. 
We find that the different energy eigenvalues  belonging to different values of
$\nu$ in a basic triangular plaquette $\alpha$ to be as shown in Table 2.
Now the lowest of these, $E_{\rm min}^{\alpha}$, will be $-U/4$ if 
\begin{equation}
-U/4\le {\rm Min}\left[ \pm t-U/12, \pm t/2-U/12, -t-U/6, t/2-U/6, 0 \right]~.
\end{equation}
This is true for $U\ge12 \mid t\mid$. Then $E_{\rm min}=-U/4$. Consequently,
$E_{\rm lo}=2N\times(-U/4)+NU/2=0=E_{\rm up}$.

So, in the $U>0$ sector exact ground state energy $E_0=0$ for $U \ge 12 \mid t
\mid$ and the ground state becomes a paramagnetic insulator with all sites
singly occupied.
\vskip .02in
%\centerline {\bf Table 2}
%\vskip .1in
\begin{center}
\begin{tabular}{|c||c|}
\hline
$\nu$ & Energy \\
\hline
0, 6  & 0\\
1, 5& $-t-U/12$, $t/2-U/12$ \\
2, 4& 0, $-t-U/6$, $t/2-U/6$ \\
3 & $\pm t-U/12$, $-U/4$, $\pm t/2-U/12$ \\
\hline
\end{tabular}
\end{center}
\vskip .01in
\noindent
{\bf Table 2:} Eigenvalues of the cluster-Hamiltonian $H_{\rm cluster}(\alpha)$
as given in (5) corresponding to different number ($\nu$) of particles.
\vskip .02in

\noindent
II. $U<0$: Here we choose the trial wavefunction as follows:
\begin{equation}
|\Psi \rangle = \prod_{i \in {\cal L}} c_{i \uparrow}^{\dag}
 c_{i \downarrow}^{\dag}~\mid \! 0 \rangle,
\end{equation}
where $\cal L$ denotes a set of arbitrarily chosen $N/2$ sites. This choice
gives $E_{\rm up}=NU/2$. Now using the possible values of the lowest energies
in a triangular plaquette as listed in Table 2 we find that $E_{\rm min}=0$ if
\begin{equation}
0\le {\rm Min}\left[ \pm t-U/12, \pm t/2-U/12, -t-U/6, t/2-U/6, -U/4 \right]~.
\end{equation}
This corresponds to $U\le -12 \mid t \mid$ and consequently $E_{\rm lo}=NU/2=E_{\rm up}$
.

So, in the $U<0$ sector exact ground state energy $E_0=NU/2$ for 
$U \le -12 \mid t 
\mid $ and the ground state becomes an insulator composed of $N/2$ ``doublon''s. 
\vskip .04in
\noindent
{\bf IV. Results obtained in the RG}
\vskip .03in
 
The ground state energy as a function of the coupling constant $U/t$, calculated
in the RG, is shown in Figs. 2(a) and 2(b)
for the two cases, $t>0$ and $t<0$ respectively. It appears that the curves
completely agree with the asymptotic  solutions given above. In fact $E_0$ 
reaches the
zero value well before the threshold value of $U$, $U_{c_1}=12 \mid t \mid$ as one increases the value of $U/t$. However,
there appears a difference between the two values of $U_{c_1}$ for $t>0$ 
and $t<0$.
This merely reflects the lack of $t \rightarrow -t$ symmetry owing to
the tripartite nature of the lattice. Also on the other side of $U/t=0$ 
 the energy
curves are in complete agreement with the available exact results. Here also
$E_0$ takes up the value $NU/2$ at a much higher value of $U$ compared to the
threshold value $U_{c_2}=-12 \mid t \mid$ as the value of $U/t$ is decreased.
It appears that the wide region in the parameter space ($-12\mid t\mid \le
U\le 12 \mid t \mid$) for which the exact
solution could not be obtained is much narrower in reality. The energy
curve obtained in the RG calculation smoothly interpolates between the
two exactly solvable opposite limits. In the intermediate region
we find that $t>0$ always gives the lower energy. This can be checked against
a naive calculation for any reasonable size of a cluster having a triangular
geometry.

The local moment, which measures the proliferation of ``doublons'' in the ground
state, is plotted against $U/t$ for $t>0$ and $t<0$ in Figs. 3(a) and 3(b) 
respectively.  As we can readily see the singly occupied insulating phase
given by the wavefunction (6) corresponds to $L_0=3/4=0.75$ for 
$U\ge 12 \mid t \mid$.
Similarly, for $U\le -12 \mid t \mid$, the wavefunction (8) corresponds to 
$L_0=0$. These are reproduced in the RG in accordance with the energy curves.
The parameter space in between these two insulating phases shows a non-trivial 
discrepancy between the cases $t>0$ and $t<0$. A large plateau at $L_0=0.5$
appears for $t>0$ which is totally absent in the case of $t<0$. For the latter 
we find  a wide plateau at $ L_0=0.25$. Other small structures in the local 
moment curves may be consequences of finite size effects. The plot of the local
moment clearly shows that the extent of pairing (in the form of doublons) is 
different in the two cases although both of them possibly correspond 
to a metallic ground state. It is important to note that such a metallic
phase in the same model on 1-D chain gives free fermionic local moment $L_0=0.375$ [6].
These plots also show that the phase transitions occurring at $U_{c_1}$ and
$U_{c_2}$ are abrupt as in the 1D case [6]. The departure from $L_0=0.375$
in the metallic case is a fall-out of the lattice geometry. To naively
 illustrate
this point let us consider the specific case of $U=\mid t\mid=1.0$.
As one can readily check from Table 2 the lowest energy in each triangular
plaquette $\alpha$ comes from $\nu=2,4$ if $t>0$; typical configurations
corresponding to this is shown in Fig. 4(a). Clearly such configurations
will dominate in the global wavefunction and the value of $L_0$ will be 
pushed towards $0.5$. On the other hand, a similar observation reveals that
in case of $t<0$ lowest contribution comes from the $\nu=3$ sector with
configurations similar to those shown in Fig. 4(b). These obviously lower
the  value of $L_0$ towards $0.25$. In reality, however, $L_0$ is slightly
greater than $0.25$ at this point for $t<0$ as one can see from Fig. 3(b). This
is because of the mixing of other configurations for optimization of the hopping
 process between different clusters. Also the finite-size effects might be 
there.

The difference in the values of
$U_{c_2}$ for $t>0$ and $t<0$ is distinctly visible from the plots of the local 
moment. Thus the local moment plot very clearly captures the lack of
$t \rightarrow -t$ symmetry which is essential for the lattice under
consideration.

\vskip .04in
\noindent
{\bf V. Conclusion}
\vskip .03in

Summarizing, we have studied the Hubbard model with the bond-charge interaction
on a triangular lattice at the special point of particle-hole
symmetry. At this point we obtain exact results in two opposite
limits of the parameter space. The system behaves as paramagnetic
insulator above a certain value $U_{c_1}$ of the on-site correlation.
Below the critical value $U_{c_2}$ it undergoes a transition to 
an insulating state of disordered doublons.
To explore the full parameter space
we employ a real space version of the RG. The RG scheme is suitably
adapted for this purpose. The ground state energy and the local moment
values calculated in the RG reproduce the exact results as it did in the 
1-D case [6] too. This lends some support for the present RG approximations. 
In the intermediate
range of the parameter space, where no exact solution has been available,         the RG results indicate that the degree of pairing (in terms of the
formation of local doublons) is different from that in the 1-D counterpart
of this problem. The triangular geometry plays a crucial role there.
Moreover, in this region, both the energy and the local moment plots
show up the effect of loosing the $t \rightarrow -t $ symmetry. The  
parameter space between $U_{c_1}$ and $U_{c_2}$, corresponding to
a possible metallic phase (as it was in the 1-D counterpart [6, 19]),
appears to be less wider in the RG calculations compared to the
exact solution. The present study gives an indication that the RG
scheme used here could be successfully used to other cases in 2-D.
Extension to the cases $X \neq t$ seems interesting although it is
well known that the lack of particle-hole symmetry creates some
problem with the present form of the RG  on a non-bipartite lattice.
It is interesting to look for the short-ranged
correlations, if any,  in the intermediate region of the parameter space. 
It is also interesting to know the effect of the finite block-size on the
satellite plateaus in the local moment plot; this requires a larger
block in the RG analysis (a body-centered hexagon is the choice
next to a triangle). However, the convergence of the RG results to the
exact ones is often slow with the increasing block-size, and the
effect of the discarded states may be important in determining the global
wavefunction [8, 20]. Therefore, a better way of supplementing the present
study is to use the DMRG algorithm, which can take into account
a larger number of configurations within a block in a controlled and systematic
way. Of course the DMRG algorithm has to be suitably adapted (on a 2-D lattice
in the thermodynamic limit) for this purpose.
%It is interesting to look for the short-ranged
%correlations, if any,  in the intermediate region of the parameter space. For
%achieving more accuracy in the results this study could be supplemented
%by a suitably adapted (for a triangular lattice in the thermodynamic
%limit) DMRG algorithm. 

\vskip .02in

\noindent
{\bf Acknowledgment}

 One of the authors (SS) acknowledges the financial support given by 
Sonderforschungsbereich 341 supported by the Deutsche Forschungsgemeinschaft 
(DFG). Computational facilities enjoyed at the Saha Institute of Nuclear
Physics, Calcutta are also gratefully acknowledged.

\vskip .04in
\noindent
{\bf Appendix}
\vskip .03in
	In the present appendix we give a brief sketch of the derivation of
the recursion relation (3) quoted in the text.
We have retained a suitable state of low energy in each of the 
subspaces $\{ \nu=2, S=S_z=0\}$, $\{ \nu=3,S=S_z=\frac{1}{2} \}$,
$\{ \nu=3, S=-S_z=\frac{1}{2} \}$ and $\{ \nu=4,S=S_z=0 \}$.
These states are designated by $\mid \! 0'\rangle$, $\mid\uparrow '\rangle $,
$\mid\downarrow '\rangle $ and $\mid\uparrow \downarrow'\rangle $ respectively.
Let the corresponding energy eigenvalues be $E_{0'}$, $E_{\uparrow'}$,
 $E_{\downarrow'}$, $E_{\uparrow\downarrow'}$ respectively.
As we have mentioned earlier the first and the fourth states are connected by 
the particle-hole
symmetry while the second and the third, by spin-reversal symmetry. 
It follows, therefore, that $E_{0'}=E_{\uparrow\downarrow'}=E_2$
and $E_{\uparrow'}=E_{\downarrow'}=E_3$.
These four states closely resemble the single-site states $\mid \! 0\rangle$, 
$\mid\uparrow \rangle $, $\mid\downarrow \rangle $ and $\mid\uparrow \downarrow\rangle$
in that the spin quantum numbers are the same and there is a one-to-one
correspondence between the total number of electrons, $\nu $, in such a state
and the occupation number of the corresponding single-site state
(they differ by two). We, therefore, identify a three-site block as
a ``renormalized'' site in the scaled lattice and the retained
states of the three-site block as the ``renormalized'' single-site
states [3, 4]. 

The intrablock Hamiltonian could be written within the subspace of the
truncated basis in terms of the new block-fermion operators 
\begin{equation}
H_o ' = \frac{1}{2} (E_2 + E_3 )
+2( E_2-E_3)
\left(\frac{1}{2} - n_{
\uparrow'}\right)\left(\frac{1}{2} - n_{ \downarrow'}\right)
\end{equation}
where the prime denotes the renormalized block-fermion operators.
Comparison of this ``renormalized'' block Hamiltonian with the single
site part of the Hamiltonian (2) leads to the renormalization formulae
for $U$ and $D$ in (3).

To obtain the interblock part of the Hamiltonian  we calculate
the matrix elements of the old fermion operators on the boundary
site $b$, $c_{b \uparrow}$ and $c_{b \uparrow} n_{b \downarrow}$, between the states we have retained. This leads to the renormalization of $t$ as follows
\begin{equation}
t' = 2 ( \lambda \lambda^* - \lambda \lambda'^* -\lambda' \lambda^*)
\end{equation}
which is equivalent to the last relation in (3). $\lambda$ and $\lambda'$
are already given in (4).

	Let us now illustrate the scheme of calculation of the renormalized 
parameters as given by Eqn. (3). We refer to a specific case with a choice
of states belonging to the $E$-representation (Table 1) from all the subspaces.
We find from the Table 1 that there could be several sets of states from the
subspaces with $\nu=2$ and $\nu=3$ i.e. with eigenvalues $E_2$ and $E_3$
respectively. For example, we pick up the case with $E_2= t-U/4$ 
and $E_3=-3U/4$.

The block-state $|0' \rangle$ (with $S_z=0$ and $\nu=2$) which is an eigenvector of the block-Hamiltonian belonging
to the eigenvalue $E_2=t-U/4$ and is simultaneously an eigenvector of the
rotation operators $R(C_{3v})$ is given by
\begin{eqnarray}
|0'\rangle &&= \frac{1}{2\sqrt{6}} \left[~ |0~\uparrow~\downarrow \rangle - 
|0~\downarrow~\uparrow \rangle - 2 |\uparrow~0~\downarrow \rangle + 
2 |\downarrow~0~\uparrow \rangle \right. \nonumber \\
&&+\left. |\uparrow~\downarrow~0 \rangle - |\downarrow~\uparrow~0 \rangle~\right]  
+\frac{i}{2 \sqrt{2}} \left[~|0~\uparrow~\downarrow \rangle - 
|0~\downarrow~\uparrow \rangle \right. \nonumber \\
&& - \left. |\uparrow~\downarrow~0 \rangle + 
|0~\downarrow~\uparrow \rangle ~\right]~.
\end{eqnarray}
Similarly, the block-state $|\uparrow' \rangle$ (with $S_z=1/2$ and $\nu=3$) which is an eigenvector of the 
block-Hamiltonian belonging
to the eigenvalue $E_3=-3U/4$ and is simultaneously an eigenvector of the
rotation operators $R(C_{3v})$ is given by
\begin{eqnarray}
|\uparrow' \rangle &&= \frac{1}{2\sqrt{3}} \left[~|\uparrow~\downarrow~\uparrow \rangle - 
2 |\uparrow~\uparrow~\downarrow \rangle +  |\downarrow~\uparrow ~\uparrow \rangle \right] \nonumber \\
&&+\frac{i}{2} \left[~|\uparrow~\downarrow~\uparrow\rangle - 
|\downarrow~\uparrow~\uparrow \rangle ~\right]~.
\end{eqnarray}
Here  we have expressed a given configuration of the triangular block by
$|\beta~\gamma~\delta \rangle$ ($\beta, \gamma, \delta = 0, \uparrow, \downarrow
~{\rm or}~ \uparrow \downarrow$) such that the configuration obeys the ordering of the 
site indices $1 \rightarrow \beta$, $2 \rightarrow \gamma$ and $3 \rightarrow \delta$
(see Fig. 1 for the site indices).

For the choice of these particular set of states,
$$
\lambda=\langle 0' | c_{1 \uparrow} | \uparrow' \rangle = -\frac{1}{2\sqrt{2}}+
\frac{i}{2 \sqrt{6}}~,
$$ 
while 
$$ \lambda'= \langle 0' | c_{1 \uparrow } n_{1 \downarrow} |\uparrow' \rangle =0~.
$$
$\lambda'$ may have non-zero value for other sets of states. Instead of using
$c_{1 \uparrow}$ and $n_{1 \uparrow}$ one could use operators belonging to site no.
2 or 3 to land up with the same result. This is because all the three sites
are ``boundary sites'' (as pointed out earlier) and, therefore, equivalent to each other.
These lead to the 
following RG equations
\begin{eqnarray}
&&U' = 2(E_2-E_3)=2t+U \nonumber \\
&&D' = 3D  + (E_2+E_3)/2 = 3D + t/2 -U/2 \nonumber \\
&&t' = 2 Re[~\lambda^{\ast} (\lambda - 2 \lambda')~]~ t = \frac{1}{6} t~.
\end{eqnarray}

The renormalized parameters could be easily calculated in a similar way for any other
choice of states shown in Table 1.

%\newpage
\vskip .04in
\noindent
{\bf References}

\noindent
[1] Lieb E  and  Wu F 1968 {\em  Phys. Rev. Lett.} {\bf 20} 1445 

\noindent
[2] Vollhardt D 1993  {\em Correlated Electron Systems} ed 
V J Emery (World Scientific) vol 9

\noindent
[3] Hirsch J E 1980 {\em Phys. Rev.} B {\bf 22} 5259 

\noindent
[4] Dasgupta C and Pfeuty P 1981 {\em J. Phys.} C {\bf 14} 717 

\noindent
[5] Bhattacharyya B and Roy G K 1995 {\em J. Phys.: Condens. Matter} {\bf 7}
5537 

\noindent
[6] Bhattacharyya B and Sil S 1995 {\em J. Phys.: Condens. Matter} {\bf 7}
6663 \\
Bhattacharyya B and Sil S 1996 {\em J. Phys.: Condens. Matter} {\bf 8} 911

\noindent
[7] Sil S and Bhattacharyya B 1997 {\em Phys. Rev.} B {\bf 54 } 14349 

\noindent
[8] White S R 1992 {\em  Phys. Rev. Lett.} {\bf 69} 2863 
White S R 1993 {\em Phys. Rev. B} {\bf 48} 10345 

\noindent
[9] Perez-Conde J and  Pfeuty P 1993 {\em Phys. Rev.} B {\bf 47} 856 \\
Vanderzande C 1985 {\em J. Phys. A: Math. Gen.} {\bf 18} 889

\noindent
[10] Machida K and Fujita M 1990 {\em Phys. Rev.} B {\bf 42} 2673 

\noindent
[11] Krishnamurthy H R,  Jayaprakash C, Sarker S and Wenzel W 1990 {Phys. Rev. 
Lett.}  {\bf 64} 950 \\
Jayaprakash C, Krishnamurthy H R, Sarker S  and Wenzel W 1991 {\em Europhys. 
Lett.} {\bf 15} 625 

\noindent
[12] Kato M and Kokubo F 1994 {\em Phys. Rev.} B {\bf 49} 8864 

\noindent
[13] Pecher Udo and B$\ddot{\rm u}$ttner H 1995 {\em  Z. Phys.} B {\bf 98} 239 

\noindent
[14] Gazza C J, Trumper A E and Ceccatto H A 1994 {\em J. Phys.: Condens. 
Matter} {\bf 6} L624 

\noindent
[15] Hubbard J 1963 {\em Proc. Roy. Soc. London} A {\bf 276} 238 

\noindent
[16] Kino H and Kotani H {\em preprint} cond-mat/9807147\\
Kondo H and Moriya T {\em preprint} cond-mat/9807322
 
\noindent
[17] Strack R and Vollhardt D 1993 {\em  Phys. Rev. Lett.} {\bf 17} 2637 

\noindent
[18] Cotton F A 1971 {\em Chemical Applications of Group Theory} (John Wiley \&
Sons, 2nd ed)

\noindent
[19] Arrachea L and  Aligia A A 1994 {\em Phys. Rev. Lett.} {\bf 73} 2240 

\noindent
[20] Bhattacharyya B and Sil S 1996 {\em Phys. Lett.} A {\bf 210} 129 

\end{document}